# MatchThem:: Matching and Weighting after Multiple Imputation

*Farhad Pishgar [a], Noah Greifer [b], Clémence Leyrat [c], Elizabeth Stuart [b,d]*

[a] *Russell H. Morgan Department of Radiology and Radiological Science, Johns Hopkins University School of Medicine, Baltimore, United States*

[b] *Department of Mental Health, Johns Hopkins Bloomberg School of Public Health, Baltimore, United States*

[c] *Department of Medical Statistics, Faculty of Epidemiology and Population Health, London School of Hygiene & Tropical Medicine, London, United Kingdom*

[d] *Department of Biostatistics, Johns Hopkins Bloomberg School of Public Health, Baltimore, United States*

Corresponding Author

**Farhad Pishgar, MD, MPH,**

The Russell H. Morgan Department of Radiology and Radiological Science,

Johns Hopkins University School of Medicine,

601 N Caroline St, JHOC 4240, Baltimore, MD 21287;

Telephone: (443)-371-1648; Fax: (410)-502-6454;

E-mail: Pishgar@JHMI.edu

**Abstract**

Balancing the distributions of the confounders across the exposure levels in an observational study through matching or weighting is an accepted method to control for confounding due to these variables when estimating the association between an exposure and outcome and to reduce the degree of dependence on certain modeling assumptions. Despite the increasing popularity in practice, these procedures cannot be immediately applied to datasets with missing values. Multiple imputation of the missing data is a popular approach to account for missing values while preserving the number of units in the dataset and accounting for the uncertainty in the missing values. However, to the best of our knowledge, there is no comprehensive matching and weighting software that can be easily implemented with multiply imputed datasets.

In this paper, we review this problem and suggest a framework to map out the matching and weighting multiply imputed datasets to 5 actions as well as the best practices to assess balance in these datasets after matching and weighting. We also illustrate these approaches using a companion package for R, `MatchThem`.

**Keywords**

Matching; Missing Data; Multiple Imputation; Propensity Score; Rubin's Rules; Weighting

# 1. Introduction

In observational studies, there is the possibility that causal inferences between an exposure and an outcome may be confounded by imbalances in the distribution of the confounders across exposure groups. Balancing the distributions of these confounders across the exposure levels in the sample through matching or weighting is an accepted method to control for these confounders, to reduce the degree of dependence on certain modeling assumptions, and to obtain a less biased estimate of the causal effect.[1]

Despite increasing popularity in practice, these procedures cannot be immediately applied to datasets with missing values. There are several solutions to address the problem of missing data in causal effect estimation, but a standard and relatively easy-to-use one is to multiply impute the missing data, which preserves the number of units in the dataset while accounting for some of the uncertainty in the missing values.[2] However, to the best of our knowledge, there is no comprehensive matching and weighting software that facilitates causal effect estimation in multiply imputed datasets.

The present paper is aimed to review the issues around matching and weighting with multiply imputed data (sections 2 and 3), to describe the steps involved in implementing best practices for these procedures (section 4), and to introduce the `MatchThem` R package (section 5), which is designed to facilitate the application of matching and weighting methods and effect estimation to multiply imputed datasets through incorporation with multiple algorithms and statistical packages.

## 1.1. Notation

Let $i = \{1, 2, 3, \dots, n\}$ index the $n$ units in a dataset, in which the causal effects of a binary exposure indicator ($z$) on a (continuous or binary) outcome indicator ($y$) in the presence of a set of potential confounders ($X = \{x_1, x_2, x_3, \dots\}$) are to be estimated (such that $z_i = 0$ indicates that unit $i$ is assigned to the control group and $z_i = 1$ indicates that the unit $i$ is assigned to the treated group) (**Figure 1A**).

Consider a situation in which the values of the some of the potential confounders or the outcome indicator are missing for a subset of units in the observed dataset. In order to account for this missingness, the missing values are multiply imputed, creating $m$ complete datasets (such that $j = \{1, 2, 3, ..., m\}$ index these $m$ imputed datasets). Here we focus on the procedures following imputation; see [3] and [4] for accessible introductions to multiple imputation for medical researchers.

### *1.2. Software Requirements*

The `MatchThem` package works with the R statistical software and programming language and can be installed within the R software (requires ≥ 3.5.0 versions) running on different platforms. `MatchThem` can be installed from the Comprehensive R Archive Network by executing the following commands in the R software console (the `MatchThem` package depends on the `MatchIt` and `WeightIt` packages; these lines will install those packages, too):

```
>       install.packages("MatchThem")
>       library(MatchThem)
```

## 2. Matching and Weighting

Matching is a technique used to improve the robustness of the causal inferences derived from parametric and non-parametric statistical models in observational studies.[1] Matching aims to control for the $X$ (potential confounders) when estimating the relationship between $z$ (exposure indicator) and $y$ (outcome indicator) by duplicating, selecting, or dropping units from the dataset in a way that the resulting control and treated groups have comparable distributions for $X$. Despite concerns about the performance of matching methods in some instances,[5] if balance is achieved across the exposure groups in the matched sample, then bias in the causal effect estimate will be reduced.

Typically, matching relies on a distance measure to pair similar units between exposure groups, who then form the resulting matched sample; a popular distance measure is the (different of) propensity score, the predicted probability of being assigned to the treated group given the potential confounders $X$. Propensity scores can be used in nearest neighbor, full, optimal, and subclassification matching [6,7], though other distance measures and matching methods can be used as well.

Weighting is another way to achieve balance and reduce bias in a causal effect estimate. Weights for each unit can be estimated so that the distribution of potential confounders is the same across the exposure groups in the weighted samples. The weights can be used in a weighted regression of the outcome on the exposure to estimate the causal effect. A common way of estimating weights is to use a function of the propensity score, a procedure known as inverse probability weighting (IPW), though there have been some developments that bypass estimating the propensity score to estimate the weights directly.[8,9]

### 2.1. Missing Data

One of the major obstacles for most matching and weighting procedures is that they cannot be performed in a straightforward way for units with missing values in $z$ or $X$ because these procedures either search control and treated groups for units with exactly the same status for $X$ or rely on the predictions from a model with $z$ as the response variable and $X$ as the covariates, which cannot be computed in the presence of missing data.

Complete-case analysis, i.e., excluding units with missing values in the potential confounders or outcome indicator, is often the default approach for handling missing data. However, complete-case analysis may not be a valid option in all instances; the assumption of missingness completely-at-random (section 3.1), which is required to justify complete-case analysis, is often violated and it is possible that dropping units with any missing values may yield a dataset with few remaining units.[10] The standard alternative to address the problem of missing data that preserves the number of units in the dataset is to multiply impute the missing values.[11]

3. **Matching and Weighting Multiply Imputed Datasets**

Given the limitations of conducting a complete-case analysis, multiply imputing missing data before applying a matching or weighting method to the dataset with missing values has become a popular alternative.

*3.1. Multiply Imputing Missing Data*

Multiple imputation refers to the procedure of substituting the missing values with a set of plausible values that reflects the uncertainty in predicting the true unobserved values, which results in $m$ imputed (filled-in) datasets.[12] Multiple imputation is justified when the mechanism behind the missingness is ignorable, i.e., given the observed data, units with missing data represent a random subset of the dataset ('missing-completely-at-random' in Rubin's language [13]) or when the probability that a value is missing relies on values of other observed variables, but not on the missing value itself or unobserved factors ('missing-at-random' in Rubin's language [13]).

Several multiple imputation methods are described in the literature and multiple statistical packages can be used to generate multiple imputations. Generally the broad framework of these methods is the same: impute the missing values to produce $m$ datasets, analyze the imputed datasets separately, and pool the results obtained in each imputed dataset (using standard combining rules) to arrive at a single estimate for the sample.[12,13]

*3.2. Matching and Weighting Multiply Imputed Datasets*

While matching and weighting methods as the tools to estimate causal effects and multiple imputation as the flexible and general way of handling missing values are well established, there has been little work examining how to combine the two methods, and there is some debate over the correct sequence of actions for pre-processing of multiply imputed datasets by matching and weighting. There are two approaches (**Figure 2**):

(1) **The within approach:** In this approach, matching or weighting is performed within each imputed dataset, using the observed and imputed covariate values, and the causal effects estimated in each of the $m$ matched or weighted datasets are pooled together.[11]

(2) **The across approach:** In this approach, propensity scores are averaged across the imputed datasets, and, using this averaged measure, matching or weighting is performed in the imputed datasets. Finally, the causal effects obtained from analyzing the matched or weighted datasets are pooled together.[14]

The across approach has been demonstrated to have inferior statistical performance as compared to the within approach in many common scenarios,[11,15] though early research favored its use.[14] In particular, the across approach seems most effective when outcomes are not used to impute the missing covariate values.[15] Although some recommend avoiding the inclusion of the outcome variable during or prior to matching and weighting with propensity scores,[16] statistical evidence favors the use of the outcome variable in multiple imputation of covariates.[11] In addition, the across method is not compatible with methods that do not rely on a single distance measure for matching or weighting; such methods include (coarsened) exact matching,[15] genetic matching,[17] and entropy balancing,[8] which are slowly growing in popularity due to their strong performance.[1]

### 3.3. Assessing Balance in Multiply Imputed Datasets

Balance refers to the degree to which the distribution of potential confounders is similar across the exposure groups. Typically, balance is assessed by computing the standardized mean difference (SMD) and Kolmogorov-Smirnov (KS) statistic for each covariate.[18,19] When these values are small, as they would be in a randomized experiment, balance is achieved and effect estimation can proceed without fear of bias due to the observed potential confounders. Balance assessment for multiply imputed dataset has not been described previously; here we discuss best practices for balance assessment.

SMDs should be computed for each covariate within each imputed dataset. Because the bias in an effect estimate is related to the mean difference of the covariates across exposure groups, and the bias in the pooled effect estimate across datasets is the average of the biases in the imputed datasets; bias can be reduced by ensuring that the average SMD for each covariate across imputed datasets is close to zero. This recommendation relies on the idea of offsetting biases: if in some datasets the bias is positive and in others the bias is negative, on average the bias may be zero. However, even if the pooled effect estimate is unbiased, lack of balance in the individual imputed datasets can reduce the precision of the pooled estimate. Therefore, SMDs should be as close to zero as possible in each imputed dataset in addition to the average SMD across imputed datasets being small. To assess balance, we recommend the following steps:

(1) Compute the SMD and KS statistic for each covariate within each imputed dataset;
(2) Compute the average of the SMDs across imputed datasets; ideally, this value should be close to zero for each covariate;
(3) Compute the average and maximum of the *absolute* SMDs for each covariate across imputed datasets. Do the same for the KS statistics. Ideally, these should be close to zero as well, though slight departures from zero may be acceptable if the values in step (2) are close to zero.

---

[1] It should be noted that the across approach described by Mitra and Reiter [14] differs slightly from that described here; in their procedure, the averaged propensity scores are used to estimate the causal effect in a single dataset consisting of just the observed exposure and outcome values, which are assumed to be non-missing. The procedure described here is in the spirit of the original method but allows for the presence of imputed outcomes and the use of imputed covariates in the effect estimation. When there is no missingness in the outcome and covariates are not used in the effect estimation, the two versions of this approach coincide.

As in datasets without missing values, the extent of balance should be assessed on interactions between covariates and their squares and cubes.[20] In addition, the balance should be reported to ensure transparency of the analysis and to justify the validity of the estimated effect to readers. This can be done using a table or a plot, such as a Love plot,[21] which summarizes this information in a visually appealing and intuitive way. All of these steps can be performed by the `cobalt` R package [21], which interfaces directly with `MatchThem`.

## 4. Suggested Workflow

The `MatchThem` R package provides several tools and functions for proper and feasible adoption of both the within and across approaches to matching and weighting with multiply imputed data. The suggested workflow for pre-processing imputed datasets with matching or weighting using the `MatchThem` R package is as follows (**Figure 3**):

(1) **Imputing the Missing Data in the Dataset:** There are several multiple imputation methods and statistical packages for this step. Currently, the `MatchThem` package supports imputed datasets generated by the `mice` and `Amelia` packages for R.[22,23]

(2) **Matching or Weighting the Imputed Datasets:** The `MatchThem` package includes functions for matching (`matchthem()`) and weighting (`weightthem()`) the multiply imputed datasets using either the within or across approaches.

(3) **Assessing Balance on the Matched or Weighted Datasets:** Use functions in the `cobalt` R package to assess balance to ensure that the resulting bias is small across imputed datasets.[21] The `bal.tab()` and `love.plot()` functions in the `cobalt` package can be used directly on the output of `matchthem()` and `weightthem()`. If balance is not achieved, step II should be repeated with different approaches or methods until it is.

(4) **Analyzing the Matched or Weighted Datasets:** Using the `with()` function from the `MatchThem` package, causal effects and their standard errors can be estimated in each of the matched or weighted imputed datasets. Robust standard errors should be used with weighting and most matching methods and are available through integration with the `survey` package.

(5) **Pooling the Causal Effect Estimates:** The `pool()` function from the package can be used to pool the obtained causal effect estimates and standard errors from each dataset using Rubin's rules.

## 5. Example

In this section, we review the suggested workflow for matching and weighting multiply imputed datasets, using an example. The research question in this context is whether osteoporosis at baseline is associated with increased odds of developing knee osteoarthritis in the follow-up or not (Figure 1B). We will use the `osteoarthritis` dataset (included in the `MatchThem` package):

```
>       library(MatchThem)
>       data("osteoarthritis")
```

The `osteoarthritis` dataset contains data on 7 characteristics (age: `AGE`, gender: `SEX`, body mass index: `BMI`, racial background: `RAC`, smoking status: `SMK`, osteoporosis at baseline: `OSP`, and knee osteoarthritis in the follow-up: `KOA`) of 2,585 individuals. The dataset contains missing data in `BMI`, `RAC`, `SMK`, `OSP`, and `KOA` variables. We assume the missing values are missing at random.

```
>       summary(osteoarthritis)
```

### 5.1. Imputing the Missing Data in the Dataset

We use the `mice` R package to impute the missing data in the `osteoarthritis` dataset (please see the `mice` package reference manual for more details about this step [22]):

```
>       library(mice)
>       imputed.datasets <- mice(osteoarthritis,
                            m = 5, maxit = 10,
                            method = c("", "", "mean", "polyreg",
                                       "logreg", "logreg", "logreg"))
```

This command will produce 5 imputed datasets and save them in the `imputed.datasets` (`mids` class) object (the `MatchThem` package also supports imputed datasets by the `Amelia` package, please see `Amelia` package reference manual for more details [23]).

### 5.2. Matching or Weighting the Imputed Datasets

#### 5.2.1. Matching the Imputed Datasets

`matchthem()` can be used to apply the within and across matching approaches and several common matching methods, including the nearest neighbor (`"nearest"`), full (`"full"`), sub-classification (`"subclass"`), optimal (`"optimal"`), exact (`"exact"`), coarsened exact (`"cem"`), and genetic (`"genetic"`) matching methods, to multiply imputed datasets (please note that only the `"nearest"`, `"full"`, `"subclass"`, and `"optimal"` matching methods are compatible with the across matching approach because other methods do not involve estimating a distance score).

In this example, we use this function to match the multiply imputed datasets, `imputed.datasets`, using all the covariates as theoretical confounders, the within matching approach, the nearest neighbor matching on the propensity score, a caliper of 5%, and the 1:2 ratio for matching (please see the package reference manual for more details):

```
>       matched.datasets <- matchthem(OSP ~ AGE + SEX + BMI + RAC + SMK,
                          imputed.datasets,
                          approach = 'within',
                          method = 'nearest',
                          caliper = 0.05,
                          ratio = 2)
Matching Observations  | dataset: #1 #2 #3 #4 #5
```

After 5 iterations, the matched datasets will be produced and saved in the `matched.datasets` object (`mimids` class). The `mimids` objects contain data of the matching procedure and the matched datasets, which can be reviewed with `summary()` and `plot()` methods (e.g. `plot(matched.datasets, n = 2)`, where n indicates the matched dataset number), which function as they do in `MatchIt`.[6]

*5.2.2. Weighting the Imputed Datasets*

`weightthem()` can be used to estimate weights of each unit using several common weighting methods, including IPW, generalized boosted modeling weights, covariate balancing propensity score weights, and entropy balancing, in multiply imputed datasets (please see the `WeightIt` package reference manual for more details [24]).

In this example, we use this function to weight the imputed datasets, `imputed.datasets`, using all the covariates as theoretical confounders, the across weighting approach, the IPW method using logistic regression propensity scores, and targeting the average treatment effect in the matched sample (ATM) estimand (which mimics the target population resulting from matching with a caliper,[25] please note that only methods that estimate a propensity score, which include the `"ps"`, `"gbm"`, `"cbps"`, and `"super"` weighting methods, are compatible with the `"across"` approach.):

```
>       weighted.datasets <- weightthem(OSP ~ AGE + SEX + BMI + RAC + SMK,
                            imputed.datasets,
                            approach = 'across', method = 'ps',
                            estimand = 'ATM')
```

```
Estimating distances   | dataset: #1 #2 #3 #4 #5
Estimating weights     | dataset: #1 #2 #3 #4 #5
```

The `weighted.datasets` object (`wimids` class) contains data of the weighting procedure and the weighted datasets. The `wimids` class objects can be reviewed with `summary()` command (e.g. `summary(weighted.datasets, n = 3)`, where `n` indicates the matched dataset number). Please note that, as in `matchthem()`, setting the `approach = 'across'`, results in adopting a slightly different across approach from the one described by Mitra and Reiter (see details in section 3.2).[14]

### 5.3. Assessing Balance on the Matched or Weighted Datasets

#### 5.3.1. Assessing Balance on the Matched Datasets

Functions of the `cobalt` package are compatible with `mimids` objects and the extent of the balance in the matched datasets of these objects can be checked with the `bal.tab()`, `bal.plot()`, and `love.plot()` commands:[21]

```
>       library(cobalt)
>       bal.tab(matched.datasets)
Balance summary across all imputations
         Type     Min.Diff.Adj Mean.Diff.Adj Max.Diff.Adj
distance Distance      0.0128        0.0138       0.0146
AGE      Contin.      -0.0294       -0.0096       0.0154
SEX_2    Binary       -0.0011        0.0007       0.0034
BMI      Contin.      -0.0307       -0.0161      -0.0101
RAC_0    Binary       -0.0011       -0.0002       0.0011
RAC_1    Binary       -0.0078       -0.0009       0.0034
RAC_2    Binary       -0.0045        0.0011       0.0101
RAC_3    Binary       -0.0011        0.0000       0.0011
SMK      Binary       -0.0157       -0.0007       0.0158
```

This information shows that the covariates (confounders) are well balanced in the osteoporosis negative and positive groups as the averaged estimated SMD for all covariates across the imputed datasets are close to zero (step II in section 3.3). We then assess the average and maximum of the absolute SMDs for each covariate across imputed datasets:

```
>       bal.tab(matched.datasets, abs = TRUE)
```

The estimated average and maximum of the absolute SMDs for covariates are close to zero, meaning that the covariates are well-balanced in the imputed datasets (step III in section 3.3).

#### 5.3.2. Assessing Balance on the Weighted Datasets

The `cobalt` package is also compatible with the `wimids` objects and `bal.tab()`, `bal.plot()`, and `love.plot()` commands can be used on these object to assess the extent of balance the datasets of the `wimids` objects after weighting:[21]

```
>       library(cobalt)
>       bal.tab(weighted.datasets)
```
```
Balance summary across all imputations
          Type Min.Diff.Adj Mean.Diff.Adj Max.Diff.Adj
distance Distance   -0.0152      -0.0115      -0.0087
AGE       Contin.   -0.0347      -0.0275      -0.0231
SEX_2     Binary    -0.0027      -0.0005       0.0002
BMI       Contin.   -0.0080      -0.0029       0.0004
RAC_0     Binary     0.0001       0.0002       0.0002
RAC_1     Binary    -0.0024       0.0001       0.0021
RAC_2     Binary    -0.0017      -0.0007       0.0014
RAC_3     Binary    -0.0010       0.0005       0.0009
SMK       Binary    -0.0111      -0.0074      -0.0031
```
```
>       bal.tab(weighted.datasets, abs = TRUE)
```

This information shows the weighting procedure resulted in a well-balanced sample both in terms of the average of the SMDs (step II in section 3.3) and average and maximum of the absolute SMDs (step III in section 3.3) across imputed datasets for covariates.

### 5.4. *Analyzing the Matched or Weighted Datasets*

#### 5.4.1. Analyzing the Matched Datasets

The causal effect within each imputed dataset can be estimated using the `with()` command (`with()` is compatible with calls to `glm()` or similar functions with the data and weights arguments unspecified (e.g., `glm(y ~ z)`) or a call to `svyglm()` or `svycoxph()` from the `survey` package with the `design` argument unspecified (e.g., `svyglm(y ~ z)`):

```
>       library(survey)
>       matched.models <- with(data = matched.datasets,
                         expr = svyglm(KOA ~ OSP, family = binomial))
```

The calculated causal effect in each matched dataset is saved in the `matched.models` object (`mimira` class). Please note that analyzing datasets matched with replacement, with ratios other than 1:1, or with calipers, as well as weighted datasets, requires estimating robust standard errors, which is done with `svyglm()` or `svycoxph()` from the `survey` package. [26] In this example, we used ratio and caliper matching, hence, we adopt the robust method for estimating the standard errors.

*5.4.2. Analyzing the Weighted Datasets*

The weighted datasets can be analyzed similarly to the methods mentioned above:

```
>       library(survey)
>       weighted.models <- with(data = weighted.datasets,
                                expr = svyglm(KOA ~ OSP, family = binomial))
```

Results are saved in the `weighted.models` object (`mimira` class). Please note, as mentioned above, that there is no need to specify weights of units in the `expr` argument. When used with `mimira` class objects, the `with()` function automatically identifies the sampling or propensity score weight of each unit and performs weighted analyses.

### 5.5. Pooling the Causal Effect Estimates

*5.5.1. Pooling the Causal Effect Estimates (Obtained from the Matched Datasets)*

The causal effect estimates can be pooled using the `pool()` function:

```
>       matched.results <- pool(matched.models)
```

The output of the `pool()` is saved in the `matched.results` object (`mimipo` class) and has method for `summary()` command:

```
>       summary(matched.results, conf.int = TRUE)
             estimate  std.error  statistic       df    p.value      2.5 %      97.5 %
(Intercept) -0.1757256 0.09005394 -1.951337  42.5096 0.0576270 -0.3573973 0.005946089
OSP1        -0.1580366 0.13364286 -1.182529 129.6850 0.2391593 -0.4224391 0.106365845
```

The reported result here shows that, our analysis did not find an association between the osteoporosis and knee osteoarthritis development in the follow-up in this sample (beta = -0.16 [-0.42 – 0.11], odds ratio = 0.85 [0.66 - 1.11]).

*5.5.2. Pooling the Causal Effect Estimates (Obtained from the Weighted Datasets)*

The causal effect estimates obtained from analyzing the weighted datasets can be pooled similar to the above method, using the `pool()` function:

```
>       weighted.results <- pool(weighted.models)
>       summary(weighted.results, conf.int = TRUE)
             estimate  std.error  statistic       df    p.value      2.5 %      97.5 %
(Intercept) -0.1696747 0.07184652 -2.361628 197.1289 0.01916994 -0.3113612 -0.02798831
OSP1        -0.1596501 0.12408788 -1.286589 222.9289 0.19957213 -0.4041854  0.08488525
```

This confirms that our analysis didn't show an association between osteoporosis and knee osteoarthritis development in this sample.

## 6. Conclusions

Matching or weighting are accepted methods to balance the distributions of the confounders across the exposure levels in an observational study. However, these procedures cannot be immediately applied to datasets with missing values. Multiple imputation of the missing data is a popular approach to account for missing values while preserving the number of units in the dataset and accounting for the uncertainty in the missing values. In this paper, we suggested a framework to map out the matching and weighting multiply imputed datasets to 5 actions as well as the best practices to assess balance in these datasets after matching and weighting.

**Figure 1. The Research Question**

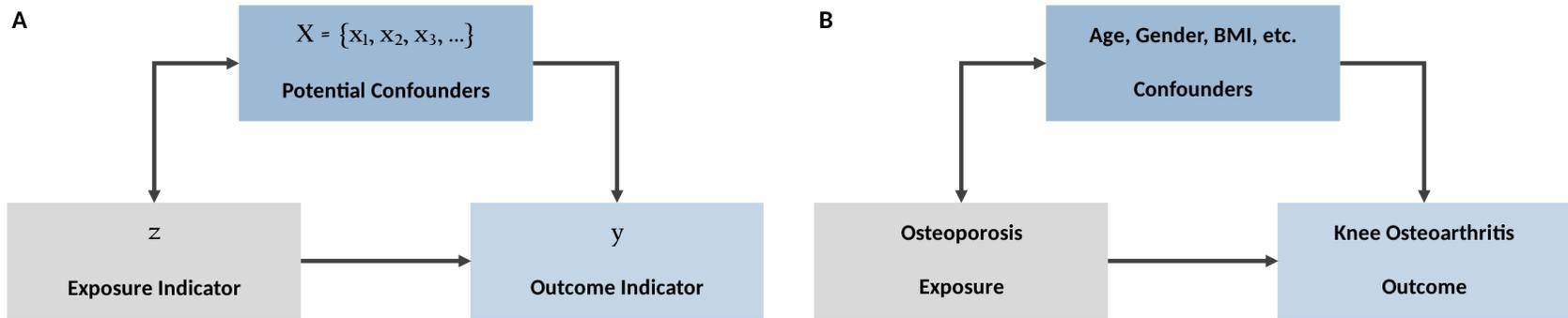

The notations used in this paper (A) and the research question used as an example in this paper (B)

**Figure 2. The Within and Across Matching Approaches**

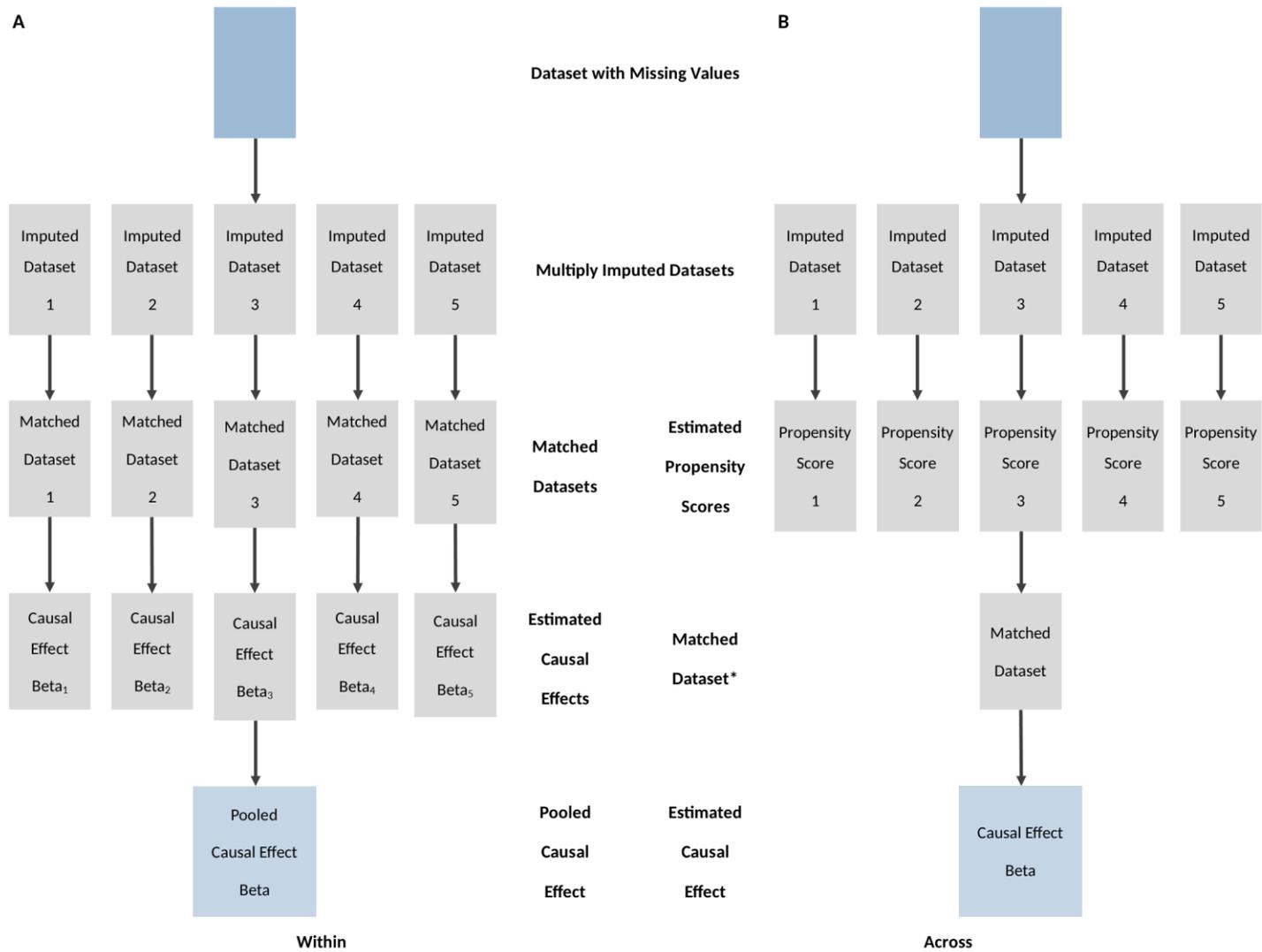

The within (A) and across (B) matching methods (please note that the `MatchThem` package uses a different across approach and performs matching in all $m$ imputed datasets, please see text for more details)

* Propensity score is averaged across datasets and the averaged measure is used for matching.

**Figure 3. Suggested Workflow for Matching and Weighting Multiply Imputed Datasets**

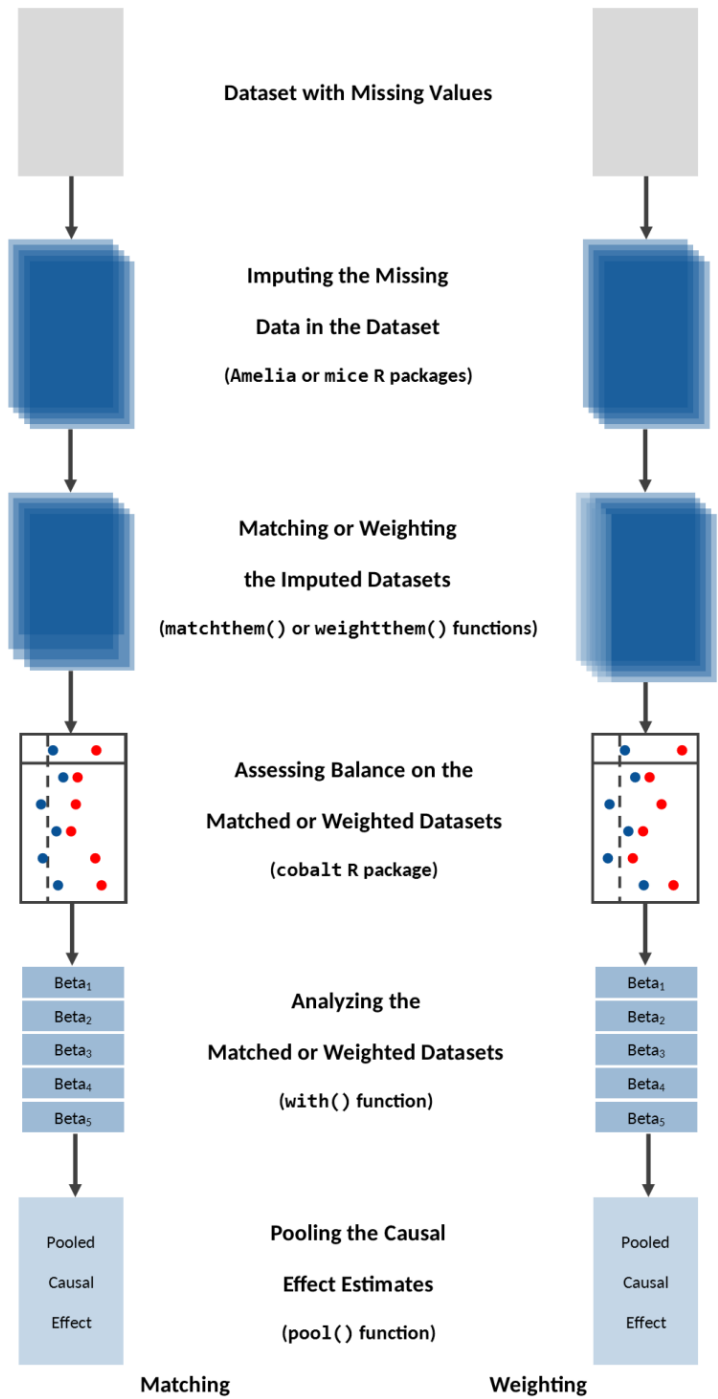